\PassOptionsToPackage{unicode}{hyperref}
\PassOptionsToPackage{hyphens}{url}
\PassOptionsToPackage{dvipsnames,svgnames,x11names}{xcolor}
\documentclass[
  a4paper,
]{article}
\usepackage{xcolor}
\usepackage{amsmath,amssymb}
\usepackage{iftex}
\ifPDFTeX
  \usepackage[T1]{fontenc}
  \usepackage[utf8]{inputenc}
  \usepackage{textcomp} 
\else 
  \usepackage{unicode-math} 
  \defaultfontfeatures{Scale=MatchLowercase}
  \defaultfontfeatures[\rmfamily]{Ligatures=TeX,Scale=1}
\fi
\usepackage{lmodern}
\ifPDFTeX\else
\fi
\IfFileExists{upquote.sty}{\usepackage{upquote}}{}
\IfFileExists{microtype.sty}{
  \usepackage[]{microtype}
  \UseMicrotypeSet[protrusion]{basicmath} 
}{}
\makeatletter
\@ifundefined{KOMAClassName}{
  \IfFileExists{parskip.sty}{%
    \usepackage{parskip}
  }{
    \setlength{\parindent}{0pt}
    \setlength{\parskip}{6pt plus 2pt minus 1pt}}
}{
  \KOMAoptions{parskip=half}}
\makeatother
\makeatletter
\ifx\paragraph\undefined\else
  \let\oldparagraph\paragraph
  \renewcommand{\paragraph}{
    \@ifstar
      \xxxParagraphStar
      \xxxParagraphNoStar
  }
  \newcommand{\xxxParagraphStar}[1]{\oldparagraph*{#1}\mbox{}}
  \newcommand{\xxxParagraphNoStar}[1]{\oldparagraph{#1}\mbox{}}
\fi
\ifx\subparagraph\undefined\else
  \let\oldsubparagraph\subparagraph
  \renewcommand{\subparagraph}{
    \@ifstar
      \xxxSubParagraphStar
      \xxxSubParagraphNoStar
  }
  \newcommand{\xxxSubParagraphStar}[1]{\oldsubparagraph*{#1}\mbox{}}
  \newcommand{\xxxSubParagraphNoStar}[1]{\oldsubparagraph{#1}\mbox{}}
\fi
\makeatother

\usepackage{longtable,booktabs,array}
\usepackage{calc} 
\usepackage{etoolbox}
\makeatletter
\patchcmd\longtable{\par}{\if@noskipsec\mbox{}\fi\par}{}{}
\makeatother
\IfFileExists{footnotehyper.sty}{\usepackage{footnotehyper}}{\usepackage{footnote}}
\makesavenoteenv{longtable}
\usepackage{graphicx}
\makeatletter
\newsavebox\pandoc@box
\newcommand*\pandocbounded[1]{
  \sbox\pandoc@box{#1}%
  \Gscale@div\@tempa{\textheight}{\dimexpr\ht\pandoc@box+\dp\pandoc@box\relax}%
  \Gscale@div\@tempb{\linewidth}{\wd\pandoc@box}%
  \ifdim\@tempb\p@<\@tempa\p@\let\@tempa\@tempb\fi
  \ifdim\@tempa\p@<\p@\scalebox{\@tempa}{\usebox\pandoc@box}%
  \else\usebox{\pandoc@box}%
  \fi%
}
\def\fps@figure{htbp}
\makeatother

\NewDocumentCommand\citeproctext{}{}

\makeatletter
 \let\@cite@ofmt\@firstofone
 \def\@biblabel#1{}
 \def\@cite#1#2{{#1\if@tempswa , #2\fi}}
\makeatother
\newlength{\cslhangindent}
\newlength{\csllabelwidth}
\newenvironment{CSLReferences}[2] 
 {\begin{list}{}{%
  \setlength{\itemindent}{0pt}
  \setlength{\leftmargin}{0pt}
  \setlength{\parsep}{0pt}
  \ifodd #1
   \setlength{\leftmargin}{\cslhangindent}
   \setlength{\itemindent}{-1\cslhangindent}
  \fi
  \setlength{\itemsep}{#2\baselineskip}}}
 {\end{list}}
\usepackage{calc}

\usepackage{amsmath}
\usepackage{amssymb}
\usepackage{amsfonts}

\usepackage{graphicx}
\usepackage{float}
\usepackage{tcolorbox} 
\usepackage{wrapfig}
\usepackage{booktabs}
\usepackage{float}

\usepackage{booktabs} 
\usepackage{multirow} 
\usepackage{multicol} 
\usepackage{makecell} 
\usepackage{siunitx}

\usepackage{hyperref} 

\usepackage[margin=1in]{geometry}
\usepackage{tabularx} 

\usepackage{fancyhdr}
\usepackage{lineno}
\usepackage[backend=biber]{biblatex}

\fancypagestyle{plain}{
  \fancyhead[L,R]{\small\color{red}PREPRINT/WORKING PAPER}
  \fancyhead[C]{\small\color{red}https://krishpn.github.io}
  \fancyfoot[L,R]{\small\color{red}PREPRINT/WORKING PAPER}
\fancyhead[C]{\small\color{red}https://krishpn.github.io}
}
\makeatletter
\@ifpackageloaded{caption}{}{\usepackage{caption}}
\AtBeginDocument{%
\ifdefined\contentsname
  \renewcommand*\contentsname{Table of contents}
\else
  \newcommand\contentsname{Table of contents}
\fi
\ifdefined\listfigurename
  \renewcommand*\listfigurename{List of Figures}
\else
  \newcommand\listfigurename{List of Figures}
\fi
\ifdefined\listtablename
  \renewcommand*\listtablename{List of Tables}
\else
  \newcommand\listtablename{List of Tables}
\fi
\ifdefined\figurename
  \renewcommand*\figurename{Figure}
\else
  \newcommand\figurename{Figure}
\fi
\ifdefined\tablename
  \renewcommand*\tablename{Table}
\else
  \newcommand\tablename{Table}
\fi
}
\@ifpackageloaded{float}{}{\usepackage{float}}
\floatstyle{ruled}
\@ifundefined{c@chapter}{\newfloat{codelisting}{h}{lop}}{\newfloat{codelisting}{h}{lop}[chapter]}
\floatname{codelisting}{Listing}

\makeatother
\makeatletter
\@ifpackageloaded{caption}{}{\usepackage{caption}}
\@ifpackageloaded{subcaption}{}{\usepackage{subcaption}}
\makeatother
\usepackage{bookmark}
\IfFileExists{xurl.sty}{\usepackage{xurl}}{} 
\hypersetup{
  pdftitle={Beyond the Numbers: Causal Effects of Financial Report Sentiment on Bank Profitability},
  pdfauthor={Krishna Neupane, Prem Sapkota, Ujjwal Prajapati},
  colorlinks=true,
  linkcolor={blue},
  filecolor={Maroon},
  citecolor={Blue},
  urlcolor={Blue},
  pdfcreator={LaTeX via pandoc}}

\title{Beyond the Numbers: Causal Effects of Financial Report Sentiment
on Bank Profitability}
\author{Krishna Neupane, Prem Sapkota, Ujjwal Prajapati}
\date{2025-06-10}
\begin{document}
\maketitle
\begin{abstract}
This study establishes the causal effects of market sentiment on firm
profitability, moving beyond traditional correlational analyses. It
leverages a causal forest machine learning methodology to control for
numerous confounding variables, enabling systematic analysis of
heterogeneity and non-linearities often overlooked. A key innovation is
the use of a pre-trained FinancialBERT to generate sentiment scores from
quarterly reports, which are then treated as causal interventions
impacting profitability dynamics like returns and volatilities.
Utilizing a comprehensive dataset from NEPSE, NRB, and individual
financial institutions, the research employs SHAP analysis to identify
influential profit predictors. A two-pronged causal analysis further
explores how sentiment's impact is conditioned by Loan Portfolio/Asset
Composition and Balance Sheet Strength/Leverage. Average Treatment
Effect analyses, combined with SHAP insights, reveal statistically
significant causal associations between certain balance sheet and
expense management variables and profitability. This advanced causal
machine learning framework significantly extends existing literature,
providing a more robust understanding of how financial sentiment truly
impacts firm performance.

\vspace{1em}

\noindent © Krishna Neupane 2025. All rights reserved. This working
paper is part of on going review for publication. No part of this
publication may be reproduced without prior permission.
\url{https://krishpn.github.io}

\vspace{1em}

 \vspace{1em}
\end{abstract}

\section{Introduction}\label{sec-emotional-introduction}

Companies' financial reports are crucial for understanding both their
historical performance and future potential, providing a wide array of
financial and non-financial data. The sentiment embedded within these
reports significantly influences key outcomes, as a firm's profitability
is deeply connected to its asset composition and capital structure. For
financial institutions, high leverage is an inherent characteristic of
their operations, not merely a strategic choice for expansion. This is
due to their core function as liquidity producers. Financial
Institutions create value by transforming illiquid, long-term assets
into highly liquid, short-term liabilities (deposits/debt). High
leverage is therefore an essential and optimal feature of their capital
structure when there is a market premium for these liquid claims (Gorton
and Winton (2017), Gorton and Winton (2003)). Therefore, the critical
focus is on how effectively this inherent leverage is managed and
capitalized to mitigate potential losses, a key aspect consistently
highlighted in financial literature.

Specifically, profitability for financial institutions critically
depends on two core aspects. First, their loan portfolio and overall
asset composition are fundamental. This composition directly determines
exposure to specific financial dangers, such as credit risk and interest
rate risk. Consequently, it significantly influences how negatively
profitability can be affected by adverse market sentiment. Broader
financial literature highlights how general investor confidence and
economic outlook constitute market sentiment, impacting firm
profitability (Financial Stability Board (2009), Robert J. Shiller
(2008), Robert J. Shiller (1992)). This study, however, takes a focused
approach. Sentiment is quantified using a classification score from
pre-trained FinancialBERT (Hazourli (2022)). This advanced machine
learning tool is applied specifically to firms' quarterly financial
reports. The second crucial aspect is balance sheet strength. This
involves the active management of inherent leverage. A firm's capital
structure---its mix of equity and debt---dictates its resilience. This
determines how well the firm handles economic downturns or shifts in
investor confidence. Institutions with insufficient capital or poor
leverage management are inherently more vulnerable to financial shocks.
Therefore, both asset composition and leverage management are vital.
They directly shape a firm's risk profile and determine its capacity for
sustainable profits amid market volatility. Against this backdrop, this
study aims to achieve two primary objectives. First, it investigates the
causal impact of financial report sentiment on firm profitability.
Second, it explores how effective leverage management and asset
composition mediate this relationship, influencing the stability and
resilience of financial institutions.

However, complex financial phenomena pose significant methodological
challenges in empirical analysis. Pervasive problems include the
over-interpretation of control factors, where undue direct causal
influence is assigned. Another issue is the unintended introduction of
bias from standard omitted variable controls, which compromises result
reliability. Inconsistencies also arise when comparing regression
coefficients across diverse models (Cenci (2024)). These problems stem
from superficially applying regression for causal inference. This
underscores the critical need for rigorous ex-ante identification and
ex-post interpretation of causal structures. These limitations are
well-established in other disciplines, for example, biomedical research
(Imbens and Rubin (2015)), insider trading (Neupane et al. (2025)),
education (Hong and Raudenbush (2006)), social networks
(Goldsmith-Pinkham and Imbens (2013), Calvó-Armengol, Patacchini, and
Zenou (2009)), and so on. Their continued prevalence in empirical
finance necessitates a renewed focus on robust ex-ante causal
identification to effectively circumvent these pitfalls.

Addressing inherent biases and enabling robust causal inferences,
especially in complex, novel domains, necessitates advanced
methodologies. One such powerful approach for covariate adjustment and
heterogeneous causal effect estimation is Causal Forests (CF).
Originally proposed by Athey and Imbens (2016) and refined follow-on
studies, for example, Wager and Athey (2018), Athey, Tibshirani, and
Wager (2019), and Nie and Wager (2021), CF provide a sophisticated
machine learning framework designed to tackle these challenges
effectively.

This study offers several significant contributions to the empirical
finance literature, particularly in the domain of financial institution
profitability and sentiment analysis. First, a novel, integrated
classification-causal analysis methodology is introduced, explicitly
designed to overcome inherent limitations of traditional empirical
finance studies. While existing literature often relies on correlational
analyses or standard regression techniques, which, as discussed,
frequently suffer from challenges like omitted variable bias,
endogeneity, and difficulties in identifying heterogeneous effects
(Cenci (2024), Anand and Arya (2024)), this research pioneers the use of
CF in this specific context to establish robust, non-linear causal
effects in the context of Nepal. This represents a substantial shift
from merely predicting outcomes to understanding their causal drivers.
Second, the methodology introduces a highly precise and context-specific
approach to sentiment quantification. Unlike many prior studies that
employ generic lexicon-based sentiment analysis, for example, Loughran
and McDonald (2020), Loughran and McDonald (2016) or simpler machine
learning models on broader news or social media data, this work uniquely
integrates a pre-trained FinancialBERT model. This powerful
transformer-based architecture is applied directly to firms quarterly
financial reports, capturing nuanced financial language and jargon to
generate a more accurate and domain-specific sentiment classification
score. This fine-grained sentiment metric allows for an unprecedented
level of detail in examining the perceptions derived from formal company
disclosures. Third, the CF framework is uniquely enhanced through a
strategic integration with XGBoost and the use of SHapley Additive
exPlanations (SHAP) values for decorrelated feature importance. While
CF's foundational strength lies in estimating heterogeneous treatment
effects, this paper augments its interpretability and robustness.
Previous applications of causal inference often struggle with
transparently identifying the covariates driving heterogeneity or suffer
from feature correlation issues impacting model stability. By leveraging
XGBoost's superior performance for tabular data (Neupane and Griva
(2025); Shwartz-Ziv and Armon (2022)) and SHAP values to decorrelate and
re-rank features (Lundberg, Erion, and Lee (2019); Neupane et al.
(2025)), this methodology provides a more granular and reliable
understanding of how specific firm characteristics mediate causal
effects. This methodological advancement addresses pervasive challenges
like bias and the over-interpretation of control factors, offering clear
``daylight'\,' from methodologies that lack such integrated
interpretability and causal robustness. Collectively, these
contributions provide a more nuanced and causally rigorous understanding
of how financial report sentiment, alongside asset composition and
leverage management, impacts the profitability and resilience of
financial institutions, laying a foundation for more effective risk
management and policy decisions.

The CF fundamentally shifts from traditional machine learning by
prioritizing the understanding of causal effects over mere prediction
accuracy. These models estimate heterogeneous treatment effects,
inherently adjusting for various covariates, which allows them to
isolate treatment impacts for specific individual characteristics. CF
addresses the core causal question: ``what if'' or ``what would'' happen
under different conditions or interventions (Imbens (2000)), by
comparing ``actual'' scenarios to ``counterfactuals'' using the notion
of ``potential outcomes'' and ``treatment'' Cunningham (2021). Despite
this, the inherent inability to simultaneously observe both potential
outcomes (factual and counterfactual) remains the ``fundamental problem
of causal inference'' (Gelman (2011), Pearl (2000), Angrist and Imbens
(1995), Holland (1986)).

The CF shares conceptual foundations with Radom Forest (RF), however,
this paper leverages CF's architectural flexibility and analytical
strengths by integrating it with the distinct ensemble method, XGBoost.
XGBoost was specifically chosen for its documented superior performance
over RF, as highlighted by Neupane and Griva (2025). In short, RF and
XGBoost are decision tree ensemble methods, but XGBoost's sequential
boosting strategy typically achieves higher accuracy by optimizing for
both bias and variance, contrasting with RF's parallel bagging approach
primarily focused on variance reduction. Furthermore, Shwartz-Ziv and
Armon (2022) found that XGBoost generally outperforms deep learning
models in performance and tuning for tabular data. This strategic
CF-XGBoost integration provides a robust avenue for analyzing downstream
interpretability, capitalizing on XGBoost's inherent model
interpretability through feature importance rankings, which can be
considered an intermediary step in empirical analysis (Athey and Imbens
(2016)).

The XGBoost offers inherent feature interpretations, their utility is
limited by being based on training data and impacted by feature
correlations, as highlighted by Neupane et al. (2025). To address this,
our analysis refines feature understanding by explicitly decorrelating
and re-ranking features in the test data using SHAP values, originally
introduced by Shapley et al. (1953). SHAP values provide a more nuanced
understanding of feature importance by estimating each feature's
marginal contribution across all possible combinations (Lundberg, Erion,
and Lee (2019)), proving particularly insightful for identifying leading
factors in highly correlated and heterogeneous trade and financial
datasets and serving as a valuable intermediate step in empirical
analysis. This refined feature importance, derived from models like
XGBoost, creates a powerful synergy when connected to CF. CF, by
estimating heterogeneous treatment effects, inherently adjusts for
various covariates to isolate treatment impacts for specific individual
characteristics and is robust in overcoming issues like transportability
and selection bias. Therefore, the precise and decorrelated feature
importance identified through SHAP analysis can inform and significantly
enhance CF's capabilities, allowing it to leverage these critical
covariates to determine how causal effects vary across different firm
characteristics or market conditions. This integration ultimately
provides a more granular understanding of causal relationships and
ensures more accurate generalization of findings beyond the original
study context, effectively moving from identifying what features are
important to understanding how these features modulate causal effects
and where those effects hold true.

The complexities of financial analysis and the limitations of
traditional empirical methods have been covered (Section
\ref{sec-emotional-introduction}). Next, details are provided on how CF
can robustly identify heterogeneous causal effects (Section
\ref{sec-method-proposed-method}). Following that, data collection and
preprocessing strategies are explored (Section
\ref{sec-data-description}) before presenting a comprehensive analysis
of the causal relationships between quarterly report sentiment and firm
profitability (Section \ref{sec-analysis}). Finally, conclusions and
future directions are offered in Section
\ref{sec-conclusions-Future-Research}.

\section{Method}\label{sec-method-proposed-method}

The effectiveness of tree-based ensemble methods is underscored by their
increasing prevalence in empirical economics/finance, often serving as a
crucial intermediate step in the analytical process (Athey and Wager
(2019)). Building on the established success of these powerful ensemble
method XGBoost, this paper implements CF, a sophisticated extension that
moves beyond mere prediction to explore the ``why'' behind observed
patterns. Here, highly ranked features are introduced as ``treatments''
within CF to specifically delve into the data and uncover irregularities
stemming from underlying causal relationships. This section will briefly
introduce XGBoost, detail feature selection via SHAP analysis, and
describe the CF methodology.

\subsection{extreme Gradient Boosting
(XGBoost)}\label{sec-method-gradient-boosting}

XGBoost, proposed by Chen and Guestrin (2016), extends the generalized
gradient boosting framework introduced by Friedman, Hastie, and
Tibshirani (2000) to efficiently handle large datasets, utilizing a
greedy learning method that supports parallelization to iteratively
update the weights of weakly created trees (base learners) through a
boosting mechanism that learns from data attributes. This repeated
re-weighting refines accuracy by improving on previously unexplored data
patterns. Specifically, prediction mistakes are grouped and fed back
into the ensemble for weighted voting (Weighted Majority Algorithm). New
base learners are constructed to be maximally correlated with the
negative gradient of the loss function, where XGBoost notably utilizes
both first and second-order gradients (Hessian) in its objective
function for more precise and robust optimization. This iterative
process continues until no further new patterns are detected. The
optimization process in XGBoost aims to minimize a regularized objective
function that XGBoost solves (see Equation
\ref{xgboost-optimization-equation} for details). This function
comprises a standard loss term measuring prediction error and a
regularization term that penalizes the complexity of the tree models,
thereby controlling overfitting. At each iteration, XGBoost adds a new
tree (\(f_t\)) that minimizes this objective function, which is
efficiently approximated using a second-order Taylor expansion. For a
detailed technical description, readers are invited to reference Neupane
and Griva (2025) and Chen and Guestrin (2016).

\begin{equation}
    L^{(t)} = \sum_{i=1}^{n} \left[g_i f_t(x_i) + \frac{1}{2} h_i f_t^2(x_i)\right] + \Omega(f_t)
    \label{xgboost-optimization-equation}
\end{equation}

where $g_i$ is the first-order gradient of the loss function with respect to the prediction from the previous step, and $h_i$ is the second-order gradient (Hessian) of the loss function with respect to that same prediction. Additionally, $f_t(x_i)$ represents the prediction of the new tree being added at iteration $t$ for data point $x_i$, while $\Omega(f_t)$ is the regularization term, which penalizes the complexity of the new tree to prevent overfitting and typically includes terms related to the number of leaves and the magnitude of the leaf scores.

\subsection{Feature Importance}\label{sec-method-feature-importance}

XGBoost provides inherent feature importance metrics, yet their utility
for nuanced interpretations is limited. These metrics, frequently based
on impurity scores like Gini impurity or entropy (Xu et al. (2014),
Duchi et al. (2008), Breiman (2001)), quantify a feature's contribution
to reducing classification error during model training. However, they
can overfit to training data and yield biased assessments, especially
with highly correlated features, making individual contributions
difficult to discern (Shalit (2020)). To address these shortcomings and
provide more robust, localized, and correlation-aware interpretations,
this paper employs SHAP as introduced in
Section~\ref{sec-emotional-introduction}. Rooted in cooperative game
theory, SHAP treats each feature as a ``player'' whose cooperation
contributes to the model's output (the ``collective gain''). The
``payoff'' for a feature represents its marginal contribution to the
model's prediction when included versus excluded from various coalitions
of other features, with higher payoffs indicating greater feature
importance. SHAP has been successfully applied across diverse domains,
including credit scoring (Al Shiam et al. (2024), Nallakaruppan et al.
(2024), Liu et al. (2024)), socio-economic impact (Bai, Lam, and Li
(2023)), natural language processing (Huang and Zhang (2024)), corporate
voting valuation (Zingales (1992)), wage bargaining (Brugemann, Gautier,
and Menzio (2019)), financial fraud detection (Lin and Gao (2022)), and
risk attribution in banking (Tarashev, Tsatsaronis, and Borio (2016)),
among others. Importantly, SHAP satisfies four key axioms for fair
attribution, ensuring efficiency (total value is fully distributed),
symmetry (equal contributions receive equal rewards), dummy
(non-contributing features receive zero payoff), and additivity (sum of
individual SHAP values equals the sum of combined value functions) (Wang
et al. (2024)).

Formally, for a model \(f\) with \(d\) input features, where features
cooperate to produce classification values, the incremental value
(marginal contribution) of a feature \(x_j\) when joining a set of
features \(S\) is represented by Equation
\ref{eq:shapley-marginalcontribution} (Kamath and Liu (2021), Lundberg,
Erion, and Lee (2019)):

\begin{equation}
    f(x_j|S) = f(S \cup \{x_j\}) - f(S),
    \label{eq:shapley-marginalcontribution}
\end{equation}

where $x_j$ is the individual feature being assessed, $f$ is the overall model's prediction output (or more specifically, the overall model's prediction function), $S$ is the set of features excluding $x_j$, $S \cup \{x_j\}$ is the set $S$ with $x_j$ added, $f(S \cup \{x_j\})$ is the model's prediction with $S$ and $x_j$, and $f(S)$ is the prediction with only $S$. This equation evaluates how much $x_j$ improves the model's predictive power when added, given features in $S$.

\subsection{Addressing the the Causal Question with Causal
Forests}\label{sec-method-causal-forests}

This research aims to determine the
\textbf{causal effects of sentiment from quarterly reports on firm profitability dynamics}.
The methodological framework and notations are adapted from Neupane et
al. (2025) and Audrino et al. (2022). Within this framework, sentiment
is conceptualized as a \textbf{treatment variable} with two distinct
values: \textbf{negative ($n$)} and \textbf{positive ($p$)}, denoted as
\(S = \{p, n\}\). For each instance \(i\) (representing a firm at a
quarterly report), two \textbf{potential outcomes} are defined:
\(Y_i(n)\) and \(Y_i(p)\), representing profitability dynamics (e.g.,
returns, volatilities) realized under postive and negative sentiment,
respectively, as determined by a pre-trained FinBERT model. A
\textbf{causal effect} for instance \(i\) is then the contrast
\(Y_i(s) - Y_i(s')\), where \(s, s' \in S\). For example, the causal
effect of positive versus negative sentiment is \(Y_i(p) - Y_i(n)\).

A fundamental challenge in causal inference is that for any given
instance \(i\), only the potential outcome corresponding to the actual,
observed sentiment (\(W_i\)) is realized. The observed outcome \(Y_i\)
is thus expressed in Equation \ref{eq-observed-outcomes}:

\begin{equation}
    Y_i = \mathbf{1}\{W_i = n\} Y_i(n) + \mathbf{1}\{W_i = p\} Y_i(p),
    \label{eq-observed-outcomes}
\end{equation}

where, $\mathbf{1}\{\cdot\}$ is the indicator function. The unobserved potential outcome for a given instance is known as the \textbf{counterfactual}.

Lets consider the \textbf{Average Treatment Effect (ATE)}, which
represents the average difference between potential outcomes across the
entire population. For a comparison between two specific sentiment
categories \(s\) and \(s'\), the ATE is conceptually defined as
\(\mathbb{E}[Y(s) - Y(s')]\). Under the assumption of
\textbf{unconfoundedness} (conditional on observed covariates, treatment
assignment is independent of potential outcomes), the ATE for comparing
positive (\(p\)) to negative (\(n\)) sentiment is estimated from
observed data as in Equation \ref{eq-ate-estimation}.

\begin{equation}
    \text{ATE} = \mathbb{E}[Y | W=p] - \mathbb{E}[Y | W=n],
    \label{eq-ate-estimation}
\end{equation}

where, $\mathbb{E}[Y | W=p]$ is the expected observed outcome under positive sentiment, and similarly for negative sentiment. 

The ATE provides a population-level average, it can be misleading by
obscuring significant subgroup-level variations in treatment effects
(Cook, Gebski, and Keech (2004), Assmann et al. (2000)). The inability
to observe counterfactuals further complicates observational studies. To
overcome the limitations of ATE and provide a deeper understanding of
how effects vary, this research estimates
\textbf{Individualized Treatment Effects (IATEs)}, also known as
\textbf{Conditional Average Treatment Effects (CATEs)}. CATE provides a
granular view of treatment effects for each instance \(i\), accounting
for its specific feature vector (\(X_i\)). The CATE for comparing
sentiment \(s\) to \(s'\) for a given set of covariates \(x\) is denoted
by \(\tau(x)\) and defined in Equation \ref{eq-cate-estimation}.

\begin{equation}
    \text{CATE}(\tau(x)) = \mathbb{E}[Y_i(s) - Y_i(s') \mid X_i = x]
    \label{eq-cate-estimation}
\end{equation}

Specifically, for positive (\(p\)) versus negative (\(n\)) sentiment,
the CATE for instance \(i\) with covariates \(X_i\) is
\(\tau(X_i) = \mathbb{E}[Y_i(p) - Y_i(n) \mid X_i]\). To estimate these
individualized effects, we employ \textbf{Causal Forests (CF)}, a
powerful non-parametric generalization of Random Forests. The objective
within a single Causal Tree is to create leaves where treatment effects
are homogeneous internally but heterogeneous between leaves. These
individual causal trees are then ensembled to construct the Causal
Forest, with CATE estimates from each tree being averaged. This approach
captures treatment effect heterogeneity by segmenting instances into
subgroups based on their features, with splits at the node level
creating subgroups with distinct feature values. Ultimately, at the
terminal leaf nodes, \(\tau(x)\) provides the estimated treatment effect
for all instances sharing that feature profile. The final CATE estimate
from the forest is given by Equation \ref{eq-froest-construction}.

\begin{equation}
    \text{CATE}_\text{forest}(X_i) = \frac{1}{B} \sum_{b=1}^B \text{CATE}_{\text{tree}_b}(X_i)
    \label{eq-froest-construction}
\end{equation}
where $B$ is the number of trees in the forest, and $\text{CATE}_{\text{tree}_b}(X_i)$ is the CATE estimate for transaction $X_i$ from tree $b$.

To estimate CATE, CF, leverages
\textbf{Augmented Inverse Probability Weighting (AIPW)} for estimating
CATE. AIPW is chosen for its robustness in handling complex data and
analyzing heterogeneous treatment effects. It addresses confounding by
weighting observations based on their \textbf{propensity scores} which
are the estimated probabilities of receiving the observed treatment
conditional on features (Rosenbaum (2023), Rosenbaum and Rubin (1983)).
AIPW is considered ``doubly robust'' because it provides consistent
estimates if either the model for the outcome or the model for the
propensity score is correctly specified.

Building a CF for robust CATE estimation involves several distinct
methodological steps. Each tree is constructed through
\textbf{bootstrap sampling} with replacement. To further reduce variance
and prevent overfitting, \textbf{subsampling} uses only a portion of the
bootstrap sample. A critical \textbf{honesty splitting} procedure then
divides the subsample into a ``splitting'' set (for tree structure) and
an ``estimation'' set (for parameter estimation within leaf nodes),
preventing overfitting. CF's \textbf{tree construction} fundamentally
differs from RF; it maximizes the squared difference in treatment
effects,
\(\Delta \tau^2 = (\tau_{\text{left}} - \tau_{\text{right}})^2\),
between child nodes, rather than minimizing outcome variance. This
identifies optimal split points where treatment effects diverge most
significantly. Ultimately, \textbf{aggregation} combines CATE estimates
from all individual trees by averaging them (as shown in Equation
\ref{eq-froest-construction}), which reduces variance and yields a
robust, precise estimate of the individual treatment effect. Through
these sophisticated steps, the average causal effect on potential
outcomes from deploying treatments, under specified constraints and
functional forms, can be precisely measured (Athey and Wager (2019)).

In summary, bolstered by robust machine learning techniques like XGBoost
for predictive modeling and SHAP for granular feature importance,
enables us to move beyond mere correlation to identify true causal
relationships. Having established the theoretical framework of potential
outcomes, the inherent challenges of causal inference, and the advanced
capabilities of CF for estimating individualized treatment effects, we
now pivot to our research's empirical phase. The next section will
detail the specific data collection, preprocessing steps, and the
precise implementation of this sophisticated causal machine learning
framework to analyze sentiment's impact on firm profitability dynamics,
effectively bridging our theoretical concepts with the practical
quantification of these causal relationships within our financial
dataset.

\section{The Data}\label{sec-data-description}

To analyze the causal links between firm-specific sentiment from
quarterly announcements and financial outcomes, we examined companies
listed on the Nepal Stock Exchange (NEPSE) between September 2013 and
December 2024. These ten firms were chosen due to the completeness and
availability of their required financial and market data. Our
comprehensive dataset was compiled from several sources: quarterly
financial reports (balance sheets and income statements) were gathered
directly from the Nepal Rastra Bank (NRB), while market asset prices and
trading data came from NEPSE, supplemented by quarterly financial data
from individual company websites. To derive sentiment scores, we applied
a pre-trained FinBERT model to the narrative sections of these quarterly
financial reports. Sentiment labels were assigned at the sentence level
and then aggregated into a single, overall positive or negative score
for each report, directly linking to profitability data. A unique ticker
symbol was used to merge these three diverse data sources for each of
the selected companies.

\subsection{\texorpdfstring{Control Covariates
(\(X\))}{Control Covariates (X)}}\label{sec-data-description-control-covariates}

The analysis considers numerous potential confounders that could
influence both sentiment and firm profitability. These include extensive
financial performance indicators (like Interest Expense and Total
Income), key balance sheet items (such as Paid-up Capital, Reserves,
Assets, and Liabilities), and detailed debt and borrowing categories
(including Interbank Borrowing, Bonds, and various deposit types). We
also account for loan portfolio characteristics (ranging from Private
Sector loans to Staff Loans) and specific asset composition details
(like Cash Balance and Government Securities).

Furthermore, we incorporate market and trade features, such as quarterly
high and low asset prices, which are well-established in financial
theory for measuring price fluctuations and have proven analytical power
in event studies and volatility spillover analyses (Garman and Klass
(1980) and Parkinson (1980), Campbell et al. (1998), and Diebold and
Yilmaz (2012).). Liability-related features (e.g., bills payable,
interbank borrowing, deposits) are crucial for defining a financial
institution's funding mix, impacting liquidity, interest rate risk,
capital structure, and overall financial stability, concepts deeply
rooted in capital structure theories (Trenca, Zapodeanu, and Cociuba
(2016)) and bank-specific models (Gropp and Heider (2010)).

Similarly, shareholder's equity components (Paid-up Capital, Share
Premium, Retained Earnings) serve as fundamental indicators of financial
health and funding structure, widely studied within capital structure,
agency, dividend irrelevance, and signaling theories. Our analysis also
delves into operating expenses (e.g., Commission Expense, Staff Expense,
Office Operating Expenses) to assess operational efficiency and
profitability, drawing on Cost Management and Efficiency Theory.
Finally, income-related features (e.g., Interest Income, Commission and
Discount, Operating and Non-Operating Income, Net Profit/Loss) are
central to understanding a firm's profitability, earnings quality, and
revenue diversification, aligning with theories on bank diversification
and broader profitability models.

\subsection{\texorpdfstring{Outcome Covariate
(\(Y\))}{Outcome Covariate (Y)}}\label{sec-data-description-outcome-covariate}

In this study, the primary outcome variable is firm profitability,
measured by Net Profit. This focus is grounded in several foundational
theories of corporate finance. Capital Structure Theories---such as
Modigliani-Miller, Trade-off, and Pecking Order---explain how a firm's
mix of debt and equity influences its value and profitability, balancing
tax advantages against risks like financial distress and information
asymmetry. Agency Theory further contextualizes Net Profit by examining
how capital structure choices can align or misalign the interests of
managers and shareholders, thereby affecting profitability and
risk-taking behavior. Additionally, the Diamond-Dybvig Model (Diamond
and Dybvig (1983)), while primarily addressing bank runs, highlights the
critical role of liquidity management in sustaining profitability. For
financial institutions, Net Profit is also shaped by Bank-Specific
Profitability Determinants which emphasizes how the composition of a
bank's loan portfolio and assets determines its exposure to credit,
interest rate, and liquidity risks, all of which directly impact
profitability.

By leveraging these theoretical frameworks, this research investigates
the causal impact of financial report sentiment, asset composition, and
leverage on firm profitability using advanced machine learning
methodologies.

\subsection{\texorpdfstring{Treatment Covariate
(\(T\))}{Treatment Covariate (T)}}\label{sec-data-description-treatment-covariates}

This study treats sentiment---specifically, Label 0 (positive) and Label
1 (negative)---as the treatment variable to assess its causal impact on
company profitability. The role of market sentiment in driving asset
prices and firm outcomes is well established in behavioral finance.
Foundational research by Robert J. Shiller (2000) and Robert J. Shiller
(1992) demonstrated that investor confidence and expectations
significantly influence asset prices and volatility. Early quantitative
approaches, such as those by Tetlock (2007) showed that media pessimism
and negative language in firm-specific news can adversely affect market
prices, liquidity, and future earnings. Lexicon-based sentiment
analysis, introduced by Loughran and McDonald (2011) and further
developed by Loughran and McDonald (2020), enabled the extraction of
sentiment from corporate financial reports, providing deeper insights
when combined with quantitative data. Studies like Garcia (2013) and
Jiang et al. (2019) further highlighted the asymmetric and predictive
effects of sentiment on stock returns and earnings.

Advancements in machine learning have greatly enhanced sentiment
analysis in finance. Surveys such as Xing, Cambria, and Welsch (2018)
and Xing, Cambria, and Zhang (2019) outlined the use of deep learning
for sentiment-aware forecasting, while resources like the Financial
Phrasebank (Malo et al. (2014)) and SemEval tasks (Atzeni, Dridi, and
Reforgiato Recupero (2017), Maia et al. (2018)) improved sentiment
detection using domain-specific language and semantic features. The
introduction of transformer-based models, notably FinBERT (Zhao et al.
(2021)), marked a significant leap in financial text mining accuracy.
The emergence of Large Language Models (LLMs) has further advanced
sentiment analysis. Studies such as Kirtac and Germano (2024) and
Rahimikia and Drinkall (2024) demonstrated that LLMs like OPT, BERT, and
FinBERT outperform traditional sentiment dictionaries and that
year-specific LLMs can mitigate lookahead bias in financial
applications. Recent research has increasingly focused on causal
inference to evaluate sentiment's impact on financial outcomes. For
example, Audrino et al. (2022) applied a modified causal forest approach
to analyze how news sentiment affects firm returns, volatility, and
trading volume, finding that sentiment effects are amplified during
adverse macroeconomic conditions. This aligns with methodologies
developed by Lechner (2018) and Athey and Wager (2019), which underpin
the causal framework used in this study.

In this research, we utilize the pre-trained FinancialBERT model to
label quarterly financial reports, leveraging its domain-specific
capabilities.

\section{Analysis}\label{sec-analysis}

\subsection{Correlation Analysis: Heatmap and Dendrogram
Insights}\label{sec-analysis-Correlation-Analysis-Heatmap-Dendrogram-Insights}

Financial data are often highly correlated. This section presents a
heatmap that visually represents the pairwise correlation matrix among
all features. In this heatmap, the intensity and direction of color in
each cell indicate the strength and type of correlation between the
corresponding row and column features. Specifically, dark purple
signifies a perfect positive correlation (+1), while dark orange denotes
a perfect negative correlation (-1). As expected, the diagonal displays
perfect positive correlation, as it represents each feature's
correlation with itself.

\begin{figure}[!htb]
    \centering
    \includegraphics[width=0.8\textwidth]{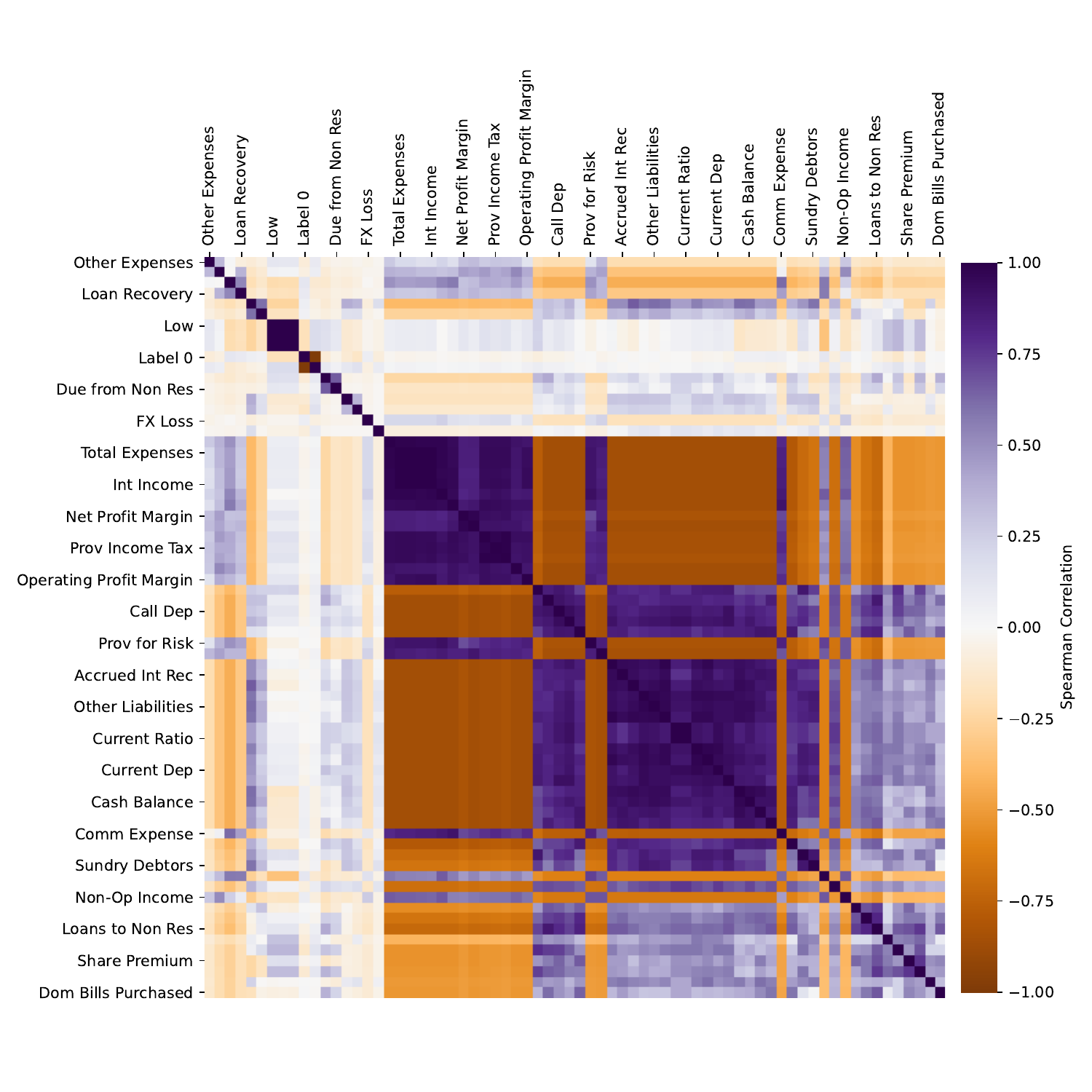}
    \caption{Heatmap: Correlation matrix showing the pairwise correlation coefficients for all features. The color scale represents the correlation value, with dark purple indicating a perfect positive correlation (+1) and dark orange indicating a perfect negative correlation (-1) }
\end{figure}

For example, we observe a smaller, distinct cluster combining Total
Expenses, Interest Income, Provision for Income Tax, and Operating
Profit Margin. This clearly represents an Income Statement and
Profitability Ratio Dimension, highlighting the expected strong
relationships among variables directly tied to a firm's financial
performance. Conversely, a larger cluster forms a Short-Term
Liquidity/Operational Balance Sheet Dimension. This group mixes features
like Accrued Interest Receivable, Other Liabilities, Current Ratio,
Current Deposits, Cash Balance, and Commission Expenses. This grouping
underscores correlations among features related to a firm's short-term
financial position, liquidity, and operational aspects, encompassing
short-term assets and liabilities, a liquidity ratio, and an operational
expense. While these clusters are valid findings from the heatmap,
indicating groups of highly correlated features, they primarily
represent distinct financial dimensions (Profitability and Short-Term
Liquidity/Operational Balance Sheet).

To uncover the inherent structural relationships among all financial
features, we employed hierarchical clustering. This method quantifies
feature similarity using a distance metric derived from one minus the
absolute pairwise correlation, with Ward's linkage criterion guiding
cluster merging to minimize the increase in total within-cluster
variance at each step. This process helps us select a representative
feature for each cluster for further analysis.

The resulting dendrogram visually represents this hierarchy,
illustrating how features progressively group based on their similarity.
At its lowest levels, the dendrogram reveals tightly clustered features
that measure very similar aspects, such as closely related financial
ratios or specific asset categories. These initial, granular clusters
then merge to form broader, more encompassing dimensions. For instance,
features related to loan portfolio and asset composition often
consolidate into a significant branch, distinct from another
representing balance sheet strength and leverage metrics. Similarly,
income statement variables tend to group into their own separate
cluster. By cutting the dendrogram at a chosen height, we can identify
these distinct feature clusters. This visualization proves invaluable
for understanding the key dimensions of financial variation,
highlighting sets of features that capture shared underlying aspects,
and ultimately informing our feature selection for subsequent analyses.

\begin{figure}[!htb]
    \centering
    \includegraphics[width=0.44\textwidth]{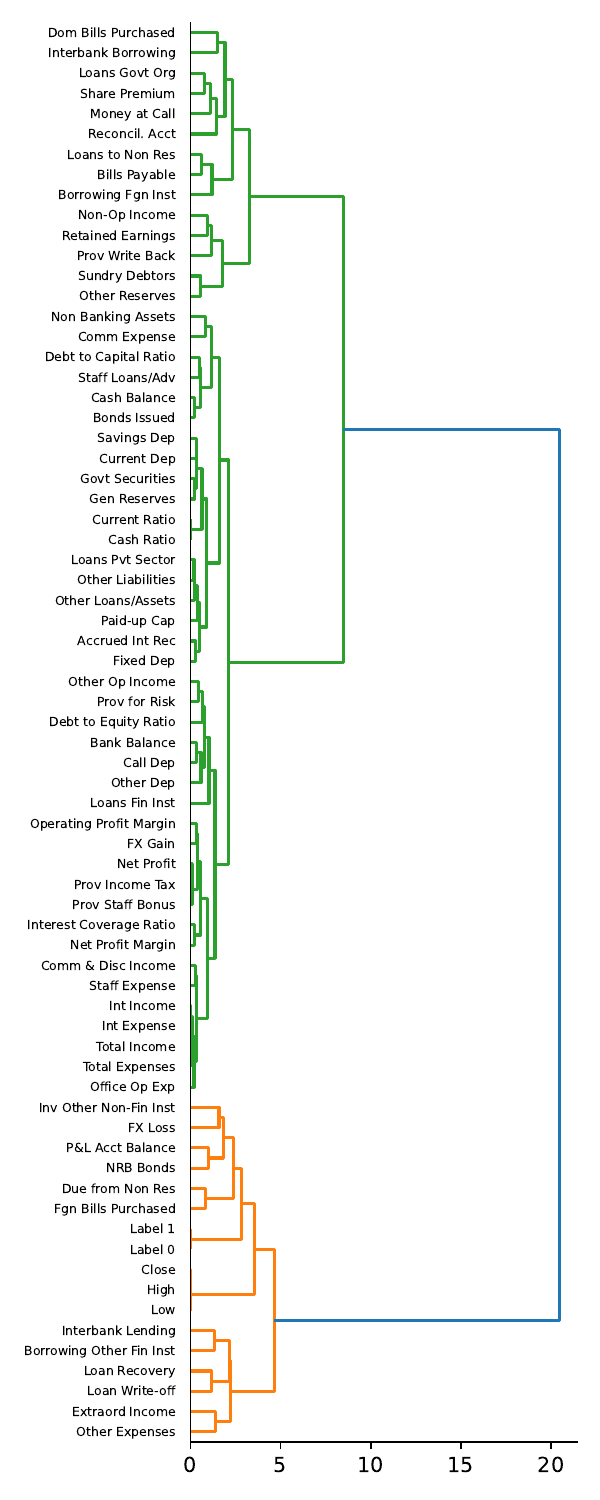}
    \caption{Dendogram: Hierarchical clustering dendrogram of financial features. The clustering, performed using Ward's linkage and a distance based on absolute correlation, visualizes the relationships and underlying dimensions among the features.}

\end{figure}

Both the dendrogram and the previously discussed heatmap are invaluable
for understanding the structural relationships among our chosen
heterogeneity features, confirming whether they capture distinct or
overlapping dimensions of a firm's financial health. Highly correlated
features within a single concept likely represent a shared underlying
factor, such as multiple ratios reflecting ``overall leverage.''
Furthermore, correlations between features spanning different financial
concepts (e.g., Asset Composition and Balance Sheet Strength) reveal how
intertwined these aspects are in our data, indicating whether
heterogeneity from asset mix is linked to that from leverage.
Identifying multicollinearity is crucial; strong correlations (exceeding
∣0.7∣ or ∣0.8∣) among intended heterogeneity factors can obscure their
unique influence on the treatment effect. In such cases, feature
selection or dimensionality reduction techniques like PCA should be
considered. Ultimately, these visualizations help us grasp the
fundamental dimensions of financial variation within our data, guiding
the strategic selection of features for our causal model to explore
heterogeneous treatment effects related to Asset Composition and Balance
Sheet Strength, thereby confirming feature relevance and illustrating
conceptual overlap.

\subsection{SHAP Analysis for Feature
Importance}\label{sec-analysis-SHAP-Analysis-for-Feature-Importance}

\begin{wrapfigure}{r}{0.60\textwidth}
    \centering
    \includegraphics[width=0.60\textwidth]{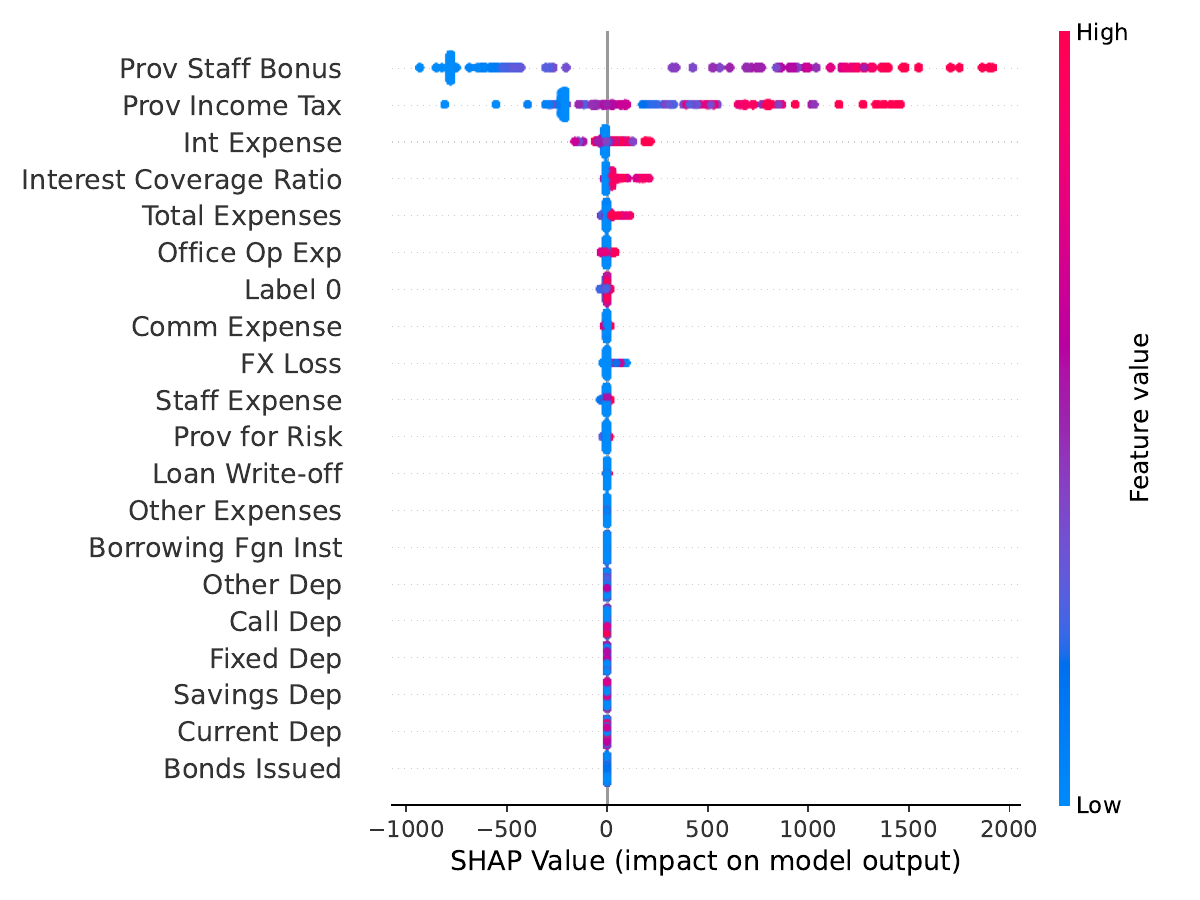 }
    \caption{SHAP Plot: SHAP beeswarm plot for the XGBoostRegressor model predicting net profit. Displays feature importance and influence on prediction, with color indicating feature value (low (blue) to high (red)).}
    \label{shapley-plot-feature-importance}
\end{wrapfigure}

While hierarchical clustering provided a robust method for initial
feature selection, we further assessed all features using XGBoost
regression (see Section \ref{sec-method-gradient-boosting}) to determine
its effectiveness in explaining profitability (see Section
\ref{sec-method-feature-importance}). To confirm the importance of these
chosen features, we performed a SHAP analysis in the test data. Unlike
the heatmap and dendrogram, which show relationships between features,
the SHAP beeswarm plot (Figure \ref{shapley-plot-feature-importance})
reveals the importance of each feature and the nature of its impact on
the target variable (predicted net profit) as learned by the model. In
the plot, features are ordered by their overall importance, with the
most impactful appearing at the top. The horizontal position of each
point indicates its SHAP value for a specific observation, showing how
much that feature's value pushed the `net profit' prediction away from
the average: positive SHAP values increase predicted `net profit', while
negative values decrease it. The color of each point signifies the
original feature value, ranging from low (blue/purple) to high (red).

The SHAP analysis of the XGBoostRegressor model, trained to predict `net
profit', reveals ``Prov Staff Bonus,'' ``Prov Income Tax,'' ``Int
Expense,'' ``Interest Coverage Ratio,'' and ``Total Expenses'' as the
most influential predictors. As anticipated, higher ``Int Expense'' and
``Total Expenses'' strongly tend to decrease predicted net profit,
consistent with their nature as costs. Similarly, a higher ``Interest
Coverage Ratio,'' indicative of stronger financial health, is
intuitively associated with higher predicted net profit. These findings
directly align with Balance Sheet Strength/Leverage and overall
profitability drivers. ``Int Expense'' and ``Total Expenses'' show the
expected inverse relationship: higher values (red) are concentrated on
the left side (negative SHAP values), indicating that increased expenses
generally lead to a lower predicted net profit. Conversely, low values
(blue/purple) for these expenses typically have little impact. The
spread of red points for ``Int Expense'' highlights variability, but the
overall negative effect of high values is clear. The ``Interest Coverage
Ratio'' aligns with financial intuition: higher values (red) are
strongly correlated with a higher predicted net profit, evident from
their concentration on the right (positive SHAP values). A higher ratio
signals stronger financial health and better ability to cover interest
costs, directly supporting increased net profit.

However, the analysis also suggests that higher values of ``Prov Staff
Bonus'' and ``Prov Income Tax'' tend to increase predicted net profit.
This initially counter-intuitive finding, given these are expenses,
likely reflects complex, indirect correlations within the dataset. It
suggests that firms with higher underlying profitability might
consequently allocate larger provisions for bonuses and taxes. This
demonstrates that models like XGBoost can identify underlying non-linear
and indirect relationships that simpler linear models might miss.
Therefore, applying interpretability techniques like SHAP is crucial not
only for confirming expected findings but also for uncovering these less
obvious patterns learned by the model. For ``Prov Staff Bonus'' and
``Prov Income Tax,'' higher values (red points) are strongly associated
with a higher predicted net profit, predominantly clustering on the
right side of the plot. Conversely, lower values (blue/purple) tend to
have less impact or are linked to lower predicted net profit. While
counter-intuitive at first glance (as these are expenses), this suggests
the model is capturing an indirect relationship where highly profitable
firms might make larger provisions for bonuses and taxes.

Similarly, for ``Office Op Exp,'' high values (red) primarily fall on
the left (negative SHAP values), meaning higher operating expenses
generally reduce predicted net profit. Low values, however, tend to
cluster around zero SHAP, indicating a negligible impact when these
expenses are minimal. Interestingly, sentiment (Label 0) shows a
different pattern. Both high (red) and low (blue) values are centered
around the mean (zero SHAP value). While there's some spread, its
overall impact on predicted net profit is smaller and less consistent
compared to the top-ranked financial features.

Overall, the SHAP results suggest that balance sheet structure (e.g.,
debt servicing ability) and overall expense levels are more predictive
of net profit in this XGBoost model than features detailing asset
composition. This indicates that, for this specific analysis, financial
structure and expense management hold greater predictive power over
asset portfolio composition in forecasting net profit.

The SHAP Beeswarm plot acted as a post-modeling interpretation tool,
explains the predictions of the trained XGBoostRegressor model. Unlike
the heatmap and dendrogram, which show relationships between features,
the SHAP plot revealed the importance of each feature and the nature of
its impact on the target variable (predicted net profit) as learned by
the model. It highlighted the most influential features for predicting
net profit and demonstrated how different values of those features push
the prediction higher or lower, including complex or non-linear
relationships not evident from simple pairwise correlations or
clustering.

From the preceding analysis (Section
\ref{sec-analysis-Correlation-Analysis-Heatmap-Dendrogram-Insights}), we
selected a representative set of features from each identified cluster
for further examination. These include: Interest Expense, Commission
Expense, FX Loss, Provision for Risk, Loan Write-off, Other Expenses,
Paid-up Capital, Share Premium, Retained Earnings, Other Reserves, Bills
Payable, Interbank Borrowing, Borrowing from Foreign Institutions,
Borrowing from Other Financial Institutions, Bonds Issued, Loans to
Financial Institutions, Domestic Bills Purchased, Foreign Bills
Purchased, Money at Call, NRB Bonds, Investments in Other Non-Financial
Institutions, Interbank Lending, High (price), Reconciliation Account,
Profit \& Loss Account Balance, Label 0 (Positive Sentiment), Provision
Write Back, Loan Recovery, and Extraordinary Income.

The features consistently identified as influential by both hierarchical
clustering and SHAP analysis include: `Comm Expense', `FX Loss', `Int
Expense', `Label 0' (Positive Sentiment), `Loan Write-off', and `Other
Expenses'. This combined list highlights a blend of financial
dimensions, rather than being exclusive to either Loan Portfolio/Asset
Composition (Asset Composition) or Balance Sheet Strength/Leverage. For
instance, `Int Expense' directly relates to debt costs, while `Loan
Write-off' is directly tied to asset quality (Asset Composition) and
indirectly impacts Balance Sheet Strength/Leverage. Other items like
`Comm Expense', `FX Loss', and `Other Expenses' are general expense/loss
items from the Income Statement, with an indirect link to Balance Sheet
Strength. This suggests that the most influential features span various
aspects of a firm's financial reporting.

\subsection{Estimation of Causal
Effects}\label{sec-analysis-Estimation-of-Causal-Effects}

Drawing on the common influential features identified in both the
correlation and dendrogram analysis (Section
\ref{sec-analysis-Correlation-Analysis-Heatmap-Dendrogram-Insights}) and
the SHAP analysis (Section
\ref{sec-analysis-SHAP-Analysis-for-Feature-Importance}), we designed an
experimental setting to validate their predictive effectiveness within a
causal framework. For this, we extracted a subset of thirteen features
with SHAP importance values greater than 0.0. These features are: `Prov
Staff Bonus', `Prov Income Tax', `Int Expense', `Interest Coverage
Ratio', `Total Expenses', `Office Op Exp', `Label 0' (Positive
Sentiment), `Comm Expense', `FX Loss', `Staff Expense', `Prov for Risk',
`Loan Write-off', and `Other Expenses'. These selected features now
serve as inputs for our causal analysis, helping us understand how
sentiment (the treatment variable) influences net profit margin (the
outcome). To thoroughly examine the impact of these top features on the
outcome, the analysis will proceed in two distinct parts in the
subsection (Section \ref{sec-analysis-Loan-Portfolio-Asset-Composition}
and Section \ref{sec-analysis-Balance-Sheet-Strength-Leverage}) .

\subsubsection{Loan Portfolio / Asset
Composition}\label{sec-analysis-Loan-Portfolio-Asset-Composition}

\begin{wrapfigure}{r}{0.7\textwidth}
    \centering
    \includegraphics[width=0.7\textwidth]{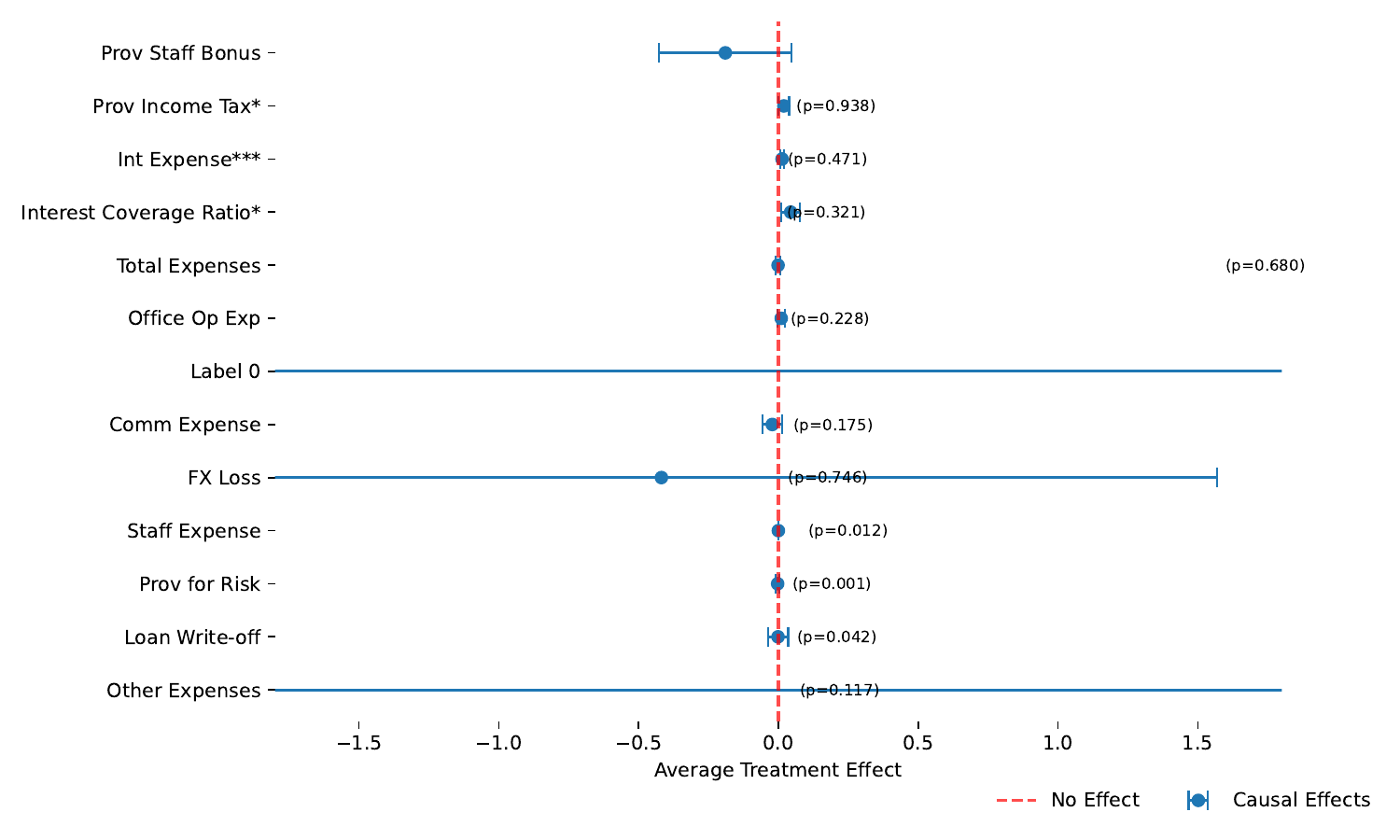}
    \caption{Average Treatment Effect plot for features in Loan Portfolio / Asset Composition Analysis. The plot shows the estimated ATE and 95 percent confidence intervals. Features are ranked as in the SHAP plot. Asterisks indicate statistical significance (* p < 0.05, ** p < 0.01  ,*** p < 0.001 ).}
    \label{ate_CausalAnalysis_Loan_Portfolio_Asset_Composition_plot}
\end{wrapfigure}

This section delves into the causal relationships between sentiment and
profitability, focusing on the role of Loan Portfolio and Asset
Composition as heterogeneous features. Beyond the thirteen top features
identified by SHAP analysis, we've introduced ``Loans Pvt Sector,''
``Loans Fin Inst,'' ``Govt Securities,'' and ``Non Banking Assets'' as
key conditioners. These variables are crucial because they explain how
sentiment's effect on profitability differs across banks, specifically
by reflecting the types of loans held or the composition of assets that
expose banks to particular risks. Essentially, these features shape the
hypothesized impact of sentiment on profitability.

Within this framework, market sentiment is our treatment, and net profit
margin is the primary Outcome variable, representing profitability. We
hypothesize that the heterogeneity in how sentiment affects
profitability is explained by variables capturing asset composition,
such as ``Loans Pvt Sector,'' ``Loans Fin Inst,'' ``Govt Securities,''
and ``Non Banking Assets.'' Furthermore, other covariates are included
to control for potential confounding factors, ensuring an unbiased
estimate of the causal effect.

This concept is well-supported by academic literature; for instance,
Hanweck and Ryu (2005) demonstrated that banks specializing in certain
types of lending or holding diverse asset mixes exhibit varying
profitability responses to shifts in interest rates or credit spreads
driven by market sentiment. This clearly illustrates how asset
composition drives heterogeneity in financial performance reactions to
external conditions.

The accompanying plot displays the estimated ATE for several financial
features, presented in the same order as the preceding SHAP beeswarm
plot to provides complementary insights into both overall importance and
individual impact heterogeneity. It includes their confidence intervals
and statistical significance levels (indicated by asterisks: * for 5
percent, *** for 1 percent). Among these features, `Int Expense' shows a
statistically significant negative ATE, with its confidence interval
entirely below zero. This aligns well with the SHAP plot, where high
values of Interest Expense consistently resulted in negative SHAP
values, indicating they decrease predicted net profit. Similarly,
`Interest Coverage Ratio' displays a statistically significant positive
ATE, with its confidence interval above zero, corresponding to the SHAP
plot where high values of `Interest Coverage Ratio' had a strong
positive impact on predicted net profit. `Prov Staff Bonus' is also
statistically significant (*), showing a slightly positive estimated
ATE. While its confidence interval visually appears to cross zero, the
asterisk indicates a significant average positive effect according to
our criteria, summarizing the varied positive impacts seen for high
values in the SHAP plot.

For most other features, including `Prov Income Tax', `Total Expenses',
`Office Op Exp', `Label 0', `Comm Expense', `FX Loss', `Staff Expense',
`Prov for Risk', `Loan Write-off', and `Other Expenses', the ATEs are
not statistically significant. This is evidenced by their confidence
intervals crossing the zero line and the absence of asterisks. This
means that although these features may show individual impact and
heterogeneity in the SHAP plot, their average causal effect across the
dataset is not statistically distinguishable from zero. It's important
to note the potential discrepancy in the provided p-values for `Staff
Expense' (p=0.012), `Prov for Risk' (p=0.001), and `Loan Write-off'
(p=0.042), which are below the significance thresholds but lack
corresponding asterisks based on your defined criteria.

In summary, the ATE plot complements the SHAP analysis by highlighting
which features have a statistically significant average impact. The SHAP
plot, conversely, provides a detailed view of how that impact varies
across individual observations and feature values. Features with
significant ATEs generally tend to be among the more important features
identified by SHAP, and the direction of the significant average effect
typically aligns with the dominant impact direction shown in the SHAP
values.

\subsection{Balance Sheet Strength /
Leverage}\label{sec-analysis-Balance-Sheet-Strength-Leverage}

\begin{wrapfigure}{r}{0.7\textwidth}
    \centering
    \includegraphics[width=0.7\textwidth]{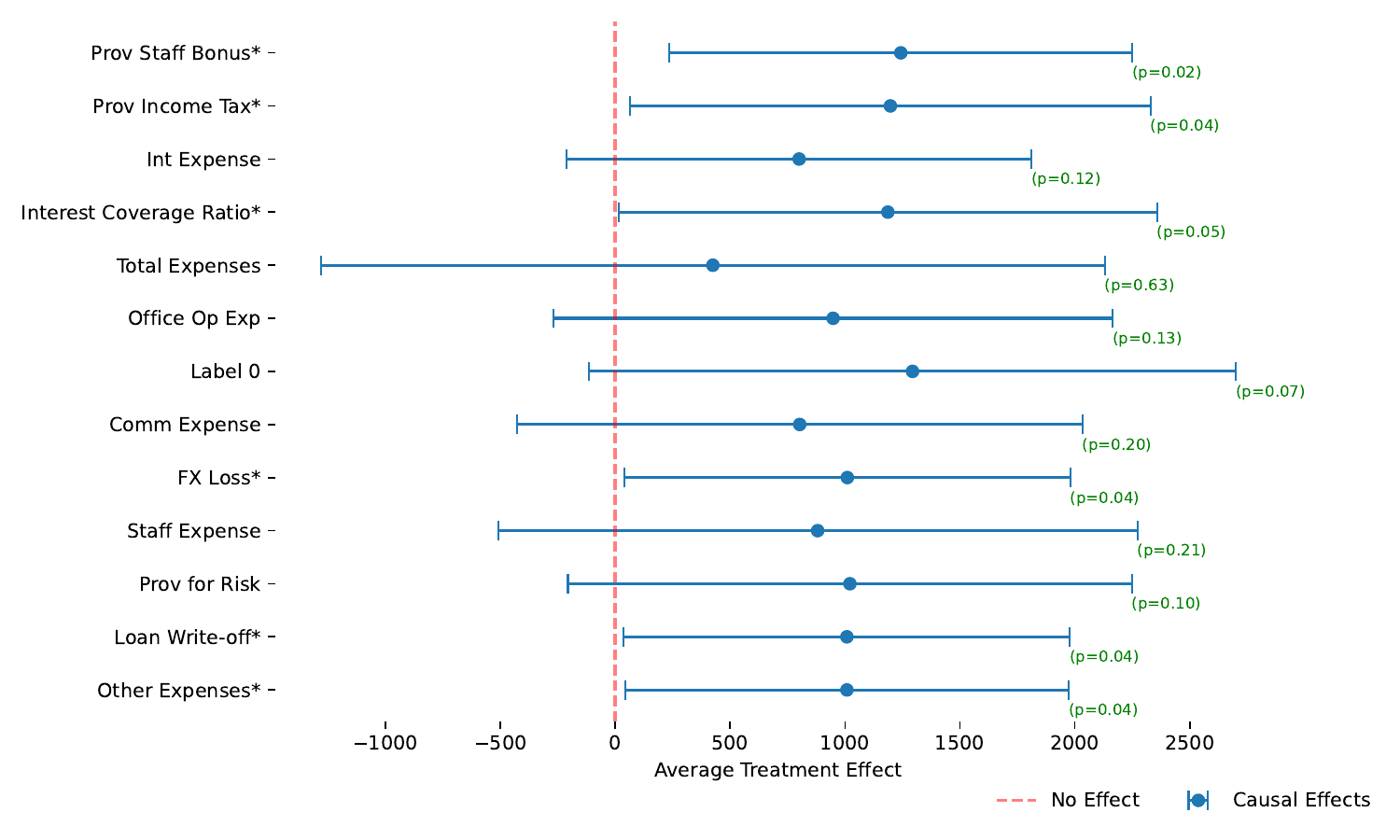}
    \caption{Average Treatment Effect plot for features in Balance Sheet Strength / Leverage Analysis. The plot shows the estimated ATE and 95 percent confidence intervals. Features are ranked as in the SHAP plot. Asterisks indicate statistical significance (* p < 0.05, ** p < 0.01  ,*** p < 0.001)}
    \label{ate_CausalForestDML_Balance_Sheet_Strength_Leverage_plot}
\end{wrapfigure}

This analysis investigates how a firm's capital structure, particularly
its debt or leverage level, influences the sensitivity of its
profitability to shifts in market sentiment, often making highly
leveraged firms more vulnerable. In this causal framework, sentiment is
the Treatment, and net profit margin is the Outcome variable. We
hypothesize that the heterogeneity in how sentiment affects
profitability is primarily driven by variables capturing leverage and
balance sheet strength. These include ``Cash Ratio,'' ``Current Ratio,''
``Debt to Equity Ratio,'' ``Interest Coverage Ratio,'' and ``Debt to
Capital Ratio''---ratios that define how sentiment's impact on
profitability varies across firms with different financial structures.
This concept is well-supported by scientific literature; for example,
Giroud and Mueller (2015) provides strong evidence that balance sheet
characteristics like leverage cause heterogeneous responses to shocks,
demonstrating how leverage can amplify a firm's sensitivity to adverse
conditions and thereby impact the sentiment-profitability link.

The plot displays the estimated Average Treatment Effects (ATEs) for
several financial features relevant to Balance Sheet Strength/Leverage
analysis, mirroring the order from the preceding SHAP beeswarm plot.

Several features exhibit statistically significant positive ATEs: `Prov
Staff Bonus' (p=0.02), `Prov Income Tax' (p=0.04), `Interest Coverage
Ratio' (p=0.05), `FX Loss' (p=0.04), `Loan Write-off' (p=0.04), and
`Other Expenses' (p=0.04). All these significant ATE estimates are
positive, with their confidence intervals entirely above the ``No
Effect'' line, indicating a robust average positive impact. This
confirms the dominant positive influence seen for higher values of
features like `Prov Staff Bonus', `Prov Income Tax', and `Interest
Coverage Ratio' in the SHAP plot. Interestingly, some features that
appeared lower in the SHAP plot, with a more mixed or near-zero average
impact, such as `FX Loss', `Loan Write-off', and `Other Expenses', show
statistically significant positive ATEs here. This suggests that within
the causal structure for balance sheet/leverage analysis, these
variables exhibit a significant average positive causal association with
the outcome, which might not be as prominent when solely looking at
their overall predictive importance or simple correlation patterns
across all predictions in the SHAP plot. Features ---`Int Expense',
`Total Expenses', `Office Op Exp', `Label 0', `Comm Expense', `Staff
Expense', `Prov for Risk'---do not have statistically significant ATEs
under this framework. Their confidence intervals cross the zero line,
indicating their average impact is not statistically distinguishable
from zero.

In summary, the ATE plot for balance sheet/leverage analysis reveals a
greater number of features with statistically significant positive
average impacts compared to the asset composition analysis. This
includes several features with lower overall importance in the SHAP
plot's ranking. This analysis suggests that within the framework of
Balance Sheet Strength/Leverage, various features---encompassing
provisions, coverage ratios, and certain expense/loss categories---show
a significant average positive association with the outcome. The
comparison with the SHAP plot highlights the distinction between overall
predictive importance/heterogeneity and statistically significant
average causal effects within a specific causal model context.

\subsubsection{Complementary Insights from SHAP and ATE
Plots}\label{complementary-insights-from-shap-and-ate-plots}

The SHAP beeswarm plot (Figure \ref{shapley-plot-feature-importance})
and the ATE plots for Asset Composition (Figure
\ref{ate_CausalAnalysis_Loan_Portfolio_Asset_Composition_plot}) and
Balance Sheet Strength/Leverage (Figure
\ref{ate_CausalForestDML_Balance_Sheet_Strength_Leverage_plot}) offer
distinct yet complementary perspectives on feature influence. The SHAP
plot provides a global view of feature importance in predicting net
profit, illustrating the heterogeneous impact of feature values across
observations. In contrast, the ATE plots estimate the average causal
effect associated with each feature within the specific causal analysis
frameworks, indicating the statistical significance of these average
effects.

\begin{itemize}
\item
  `Prov Staff Bonus': Highly important in the SHAP plot, with high
  values increasing predicted net profit. It consistently shows a
  statistically significant positive ATE in both the Asset Composition
  and Balance Sheet Strength/Leverage analyses. This suggests that, on
  average, `Prov Staff Bonus' is associated with a positive causal
  effect within both frameworks, despite its heterogeneous impact shown
  by SHAP.
\item
  `Prov Income Tax': Important in the SHAP plot, with high values
  increasing predicted net profit. While not statistically significant
  in the Asset Composition ATE analysis, it exhibits a statistically
  significant positive ATE (∗) in the Balance Sheet Strength/Leverage
  analysis. This indicates its average causal effect is only significant
  within the latter framework.
\item
  `Int Expense': Highly important in the SHAP plot, with high values
  strongly decreasing predicted net profit. It shows a statistically
  significant negative ATE (∗∗∗) in the Asset Composition analysis but
  is not significant in the Balance Sheet Strength/Leverage analysis.
  This highlights a key difference: its average negative causal effect
  is significant when considering heterogeneity related to Asset
  Composition, but not when focusing on Balance Sheet Strength/Leverage.
\item
  `Interest Coverage Ratio': Important in the SHAP plot, with high
  values increasing predicted net profit. It consistently shows a
  statistically significant positive ATE in both the Asset Composition
  and Balance Sheet Strength/Leverage analyses, indicating a robust
  positive average causal effect across both frameworks.
\item
  `Total Expenses': Important in the SHAP plot, with high values
  decreasing predicted net profit. However, it is not statistically
  significant in either the Asset Composition or Balance Sheet
  Strength/Leverage ATE analyses. This suggests that while overall
  expenses are important for predicting net profit, their average causal
  effect within these specific causal frameworks is not statistically
  significant.
\item
  `FX Loss', `Loan Write-off', `Other Expenses': These features are less
  prominent in the overall SHAP ranking but show some heterogeneous
  impact. They are not statistically significant in the Asset
  Composition ATE analysis. However, all three show a statistically
  significant positive ATE (∗) in the Balance Sheet Strength/Leverage
  analysis. This is a notable difference, suggesting that within the
  Balance Sheet Strength/Leverage causal framework, these
  loss/expense-related features have a significant average positive
  association with the outcome not captured by their overall predictive
  importance in the SHAP plot or their average effect under Asset
  Composition.
\end{itemize}

In summary, the SHAP plot provides a global view of predictive
importance and heterogeneous impact. The ATE plots, conversely, offer
specific measures of statistically significant average causal effects
under different theoretical frameworks (Asset Composition vs.~Balance
Sheet Strength/Leverage). Some features, like `Prov Staff Bonus' and
`Interest Coverage Ratio', show consistent significant ATEs across both
concepts, aligning with their SHAP importance. Others, like `Int
Expense', have a significant ATE under one concept but not the other.
Furthermore, some features less prominent in the overall SHAP ranking
gain statistical significance in the Balance Sheet Strength/Leverage ATE
analysis, highlighting that their role as heterogeneity factors can be
significant even if their overall predictive importance is lower. These
plots collectively demonstrate that feature importance, heterogeneous
impact, and statistically significant average causal effects are related
but distinct concepts, and their findings can vary depending on the
analytical framework.

\section{Conclusion and Future
Research}\label{sec-conclusions-Future-Research}

Our study thoroughly investigated the causal impact of financial report
sentiment on firm profitability using a multi-stage approach. We began
by employing \textbf{correlation analysis} (via heatmap) and
\textbf{hierarchical clustering} (via dendrogram) to understand feature
relationships and identify distinct financial dimensions, such as
profitability and liquidity. Subsequently, \textbf{SHAP analysis} on an
XGBoost model revealed key predictors of net profit, including ``Prov
Staff Bonus'' and ``Interest Coverage Ratio,'' demonstrating the model's
ability to capture complex, at times counter-intuitive, relationships.
We then established a robust causal framework using CF methods.
Combining SHAP and ATE plots across both Loan Portfolio/Asset
Composition and Balance Sheet Strength/Leverage frameworks, confirmed
that several features significantly influence how sentiment affects
profitability. This revealed nuanced insights: some features
consistently showed significant average causal effects (e.g., `Prov
Staff Bonus', `Interest Coverage Ratio'), while others exhibited
context-dependent impacts (e.g., `Int Expense') or gained significance
as heterogeneity factors despite lower overall predictive importance
(e.g., `FX Loss' in Balance Sheet Strength/Leverage). This integrated
methodology provided a robust, granular understanding of the causal
links between sentiment and profitability, moving beyond simple
correlations and underscoring the necessity of a multifaceted analytical
strategy in causal inference. Future research could extend this work by
exploring dynamic causal effects, investigating multi-valued or
continuous sentiment treatments, incorporating time-varying confounders,
and analyzing heterogeneity across different economic regimes or market
volatility.

\section*{Compliance with Ethical
Standards}\label{compliance-with-ethical-standards}
\addcontentsline{toc}{section}{Compliance with Ethical Standards}

Funding: This research received no external funding or financial
assistance during its preparation.

Competing Interests: The author certify that they have no conflicts of
interest, financial or otherwise, to disclose.

Author's Declaration on AI Assistance: The authors bear sole
responsibility for all substantive ideas and analyses within this
manuscript. Portions of the text were reviewed for language, style, and
clarity through AI-assisted copy editing, specifically using a large
language model (LLM). No autonomous content creation was performed by
the LLM.

\section*{References}\label{references}
\addcontentsline{toc}{section}{References}

\phantomsection\label{refs}

\end{document}